\documentclass[10pt,journal,compsoc]{IEEEtran}

\ifCLASSOPTIONcompsoc
  \usepackage[nocompress]{cite}
\else
  \usepackage{cite}
\fi

\usepackage{cite}
\usepackage{amsmath,amssymb,amsfonts}
\usepackage{algorithmic}
\usepackage{graphicx}
\usepackage{textcomp}
\usepackage{subcaption}

\usepackage[most]{tcolorbox}
\usepackage{fancyvrb}  
\usetikzlibrary{matrix}
\usepackage{fvextra}
\usepackage{xcolor}

\def\BibTeX{{\rm B\kern-.05em{\sc i\kern-.025em b}\kern-.08em
    T\kern-.1667em\lower.7ex\hbox{E}\kern-.125emX}}

\tcbuselibrary{breakable,skins} 

\tcbset{
  promptstyle/.style={
    breakable,
    colback=blue!5!white,  
    colframe=blue!75!black, 
    fonttitle=\bfseries,    
    fontupper=\small,
    boxrule=0.3mm,          
    arc=1mm,                
    left=1.5mm, right=1.5mm,    
    top=1.5mm, bottom=1.5mm,     
    before=\vspace{4mm} 
  }
}

\DefineVerbatimEnvironment{MyVerbatim}{Verbatim}{breaklines=true}

\usepackage{tabularx}
\usepackage{adjustbox}
\usepackage{booktabs}
\usepackage{array}

\PassOptionsToPackage{hyphens}{url}
\usepackage{hyperref}

\newcommand{\rqii}[1]{How are criteria scores of issues associated with the merge rate of Copilot pull requests?}
\newcommand{\rqiii}[1]{To what extent can we predict the likelihood of a GitHub issue resulting in a merged pull request?}
\newcommand{\rqiv}[1]{What are the most important features for predicting the likelihood of a GitHub issue resulting in a merged pull request?}

\definecolor{custom-gray}{cmyk}{0, 0, 0, 0.7, 1.00}
\newtcbtheorem[no counter]{Summary}{\hskip-0.97em}{enhanced,drop shadow={black!50!white},
  coltitle=white,
  top=0.15in,
  attach boxed title to top left=
  {xshift=1.5em,yshift=-\tcboxedtitleheight/2},
  boxed title style={size=small,colback=custom-gray}
}{summary}

\usepackage{amsmath}
\usepackage{array}

\usepackage{multicol}

\begin{document}

\title{What Makes a GitHub Issue Ready for Copilot?}

\author{Mohammed Sayagh
\IEEEcompsocitemizethanks{\IEEEcompsocthanksitem Mohammed Sayagh is with École de Technologie Supérieure, Université de Québec, Canada.\protect\\
E-mail: mohammed.sayagh@etsmtl.ca
}}

\IEEEtitleabstractindextext{%
\begin{abstract}
Large language AI-agents are extensively adopted to help developers in different coding tasks, such as developing new features, fixing bugs, as well as other software development activities like reviewing code. The AI agents are added as team members that can be integrated into the process. Developers can write a Github issue about a bug to fix or a feature to implement, then assign the issue to an AI-agent like GitHub Copilot for implementation. Based on the issue and its related discussion, the AI-agent navigates the project, performs a plan for the implementation, and executes it. Developers can then review the change either manually or through another AI-agent. However, the performance of the AI-agent and LLMs in general heavily depends on the input they receive, which is the task to perform. For instance, a GitHub issue that is unclear or not well scoped might not lead to a successful implementation that will eventually be merged. GitHub co-pilot provides a set of best practice recommendations that are limited and high-level. In this paper, we build a set of 32 detailed criteria that we leverage to measure the quality of GitHub issues to make them suitable for AI-agents. We compare the GitHub issues that lead to a merged pull request versus closed pull request. Then, we build an interpretable machine learning model to predict the likelihood of a GitHub issue resulting in a merged pull request. We observe that pull requests that end up being merged are those originating from issues that are shorter, well scoped, with clear guidance and hints about the relevant artifacts for an issue, and with guidance on how to perform the implementation. Issues with external references including configuration, context setup, dependencies or external APIs are associated with lower merge rates. We built an interpretable machine learning model to help users identify how to improve a GitHub issue to increase the chances of the issue resulting in a merged pull request by Copilot. Our model has a median AUC, precision and recall of 72\%, 69.1\% and 77.4\%, respectively. Our results shed light on quality metrics relevant for writing GitHub issues and motivate future studies further investigate the writing of GitHub issues as a first-class software engineering activity in the era of AI-teammates. 

\end{abstract}

}

\maketitle



\section{Introduction}
\label{intro}

Large language models (LLMs) are now widely used for code generation and are increasingly integrated within coding environments such as GitHub and IDEs like VS Code. LLMs are changing the development paradigm by introducing novel techniques such as vibe coding~\footnote{\url{https://cloud.google.com/discover/what-is-vibe-coding?hl=en}}, which consist of defining natural language prompts about the requirements or tasks and AI-agents to generate the code. 

For instance, GitHub Copilot can be integrated into traditional development workflows such that developers only need to describe their needs in a GitHub issue for it to be assigned to an AI-agent. Users can report a bug or request a new feature using GitHub issues, which may be enriched with additional context. 
Copilot then opens a pull request, defines a plan of action, and performs the implementation, which 
is reviewed 
before being merged. 

AI-agents heavily depend on their given prompt to 
increase solution quality. In the context of Copilot, 
the GitHub issue and its comments 
are considered input for the AI-agent, for which GitHub Copilot suggests best practices~\footnote{\url{https://docs.github.com/en/copilot/tutorials/coding-agent/get-the-best-results}}. They include clear problem descriptions, hints about files to change, and explicit validation criteria.

Writing a low-quality issue increases the likelihood of producing a pull request that diverges from the user’s intent, resulting in wasted effort. AI agents do not merely generate code; they traverse the codebase, execute commands, and run tests, all of which incur non-trivial computational costs 
without a usable outcome. Low-quality issues also waste human effort, as reviewers must analyze large, misaligned code changes that are rejected. Repeated rejections further reduce trust in AI agents, limiting their adoption and reducing their ability to be fully autonomous.

Writing good quality issues is beyond respecting a simple prompt engineering technique. A prior work~\cite{10.1145/3771933} found that advanced AI-agents need clearer instructions and guidance rather than traditional prompt engineering techniques (e.g., Zero-shot, few-shot, chain-of-thought) 
studied in prior work~\cite{white2023prompt,10795375,della2025prompt,10.1145/3722108,khojah2025impact,fagadau2024analyzing}. While a large body of research has explored which prompt-engineering techniques fit different programming tasks, prompt engineering techniques are becoming less relevant for advanced models with reasoning capabilities~\cite{10.1145/3771933}, including autonomous AI-agents that can be integrated into the GitHub workflow, such as Copilot.

In this paper, we investigate the quality of GitHub issues as inputs to AI agents and how this quality influences the likelihood of the resulting pull requests to be merged. Inspired by the GitHub best practices~\footnote{\url{https://docs.GitHub.com/en/copilot/tutorials/coding-agent/get-the-best-results}}, we define 32 issue-quality criteria (e.g., problem definition clarity, organization of the issue, and scope of the task to implement) based on which we to compare issues associated with merged versus closed pull requests. We also evaluate an \textbf{interpretable machine learning model} for predicting the likelihood of a GitHub issue leading to a merged pull request. Our approach uses LLMs with self-consistency~\cite{wang2022self} to score an issue based on the 32 criteria and leverages these scores for prediction. The goal is to assist practitioners in writing high-quality issues before affecting them to an AI-agent and consuming AI-agent tokens and human reviewer efforts. Our approach will also improve the successful integration of agents in software development.  


We first evaluate the LLMs (GPT-5-mini in this paper) capabilities on measuring the quality of an issue through the following preliminary research questions (PQ): 

\textbf{PQ1. How consistent are LLM-based evaluations of GitHub issues across repeated runs?} This research question identifies the agreement among three runs of GPT-5-mini with high reasoning capabilities. The three runs have strong, moderate, and poor agreement scores (Krippendorff) for 23, 6, and 3 out of the 32 criteria. Despite the disagreements, the difference occurs with only one of the three models (two agree on the same score, and the third provides a different score), and with only ± 1 (each criterion is scored on a scale between 0 and 5). An example of disagreement is when two agents predict a score of 2 for a given critierion and issue, while the third LLM predicts a score of 3.

\textbf{PQ2. How accurately can the model distinguish low-quality from high-quality issues?} This research question manually validates the output of the self-consistency approach (using votes) on the three runs of GPT-5-mini. We select, for each criteria, 5 issues that are predicted to be of high quality (i.e., a score of 4 or 5) and another 5 as low-quality (i.e., a score between 0 and 3) from the LLM. We manually validate, without looking at the LLM prediction, whether we can manually reach the same groups of high- and low quality issues. Our LLM based approach has a high precision of 88.75\%.

Based on the good quality of the LLM prediction, we 
aim to understand the differences between issues resulting in a merged pull request and those resulting in a closed pull request (RQ1). We further leverage the 32 criteria with other contextual features to predict the likelihood of a resulting pull request being merged from the source (i.e., the issue itself), which we evaluate in RQ2 and RQ3. 

\textbf{RQ1. \rqii} We observe that issues that are well scoped, self-contained, avoid ambiguities, provide clear guidance on the relevant artifacts related to the issue, offer implementation guidance, and break down a task into manageable steps are associated with a higher merge rate. Issues that mention external context, such as configuration, environment, context setup, external dependencies, and APIs, are associated with lower merge rates.

\textbf{RQ2. \rqiii} We find that the random forest shows the highest prediction performance, with a median AUC of 72\%, a median precision of 69.1\%, and a median recall of 77.4\%.

\textbf{RQ3. \rqiv} To increase the probability of an issue resulting in a merged pull request, our analysis of the most important features indicates that developers need to make the description shorter, well-scoped, and provide guidance on implementation. Similar to RQ1, we observe that referencing external artifacts and addressing issues that anticipate risks is associated with a lower probability of the resulting pull request being merged. 
    
\textbf{Take-home message:} Our findings indicate a strong association between the quality of GitHub issues and the likelihood of the resulting pull request being merged. \textbf{For practitioners}, we suggest writing issues that have a limited and well-defined scope, are concise and self-contained, explicitly decompose a task into manageable steps, and provide clear guidance about the implementation. GitHub issues that mention external artifacts are less likely to be merged and highlight the need for further research on how to incorporate external contexts. To assist practitioners in writing GitHub issues that are more likely to lead to a merged pull request, we propose an interpretable machine learning model that identifies which aspects of an issue should be improved. \textbf{For researchers}, our results shed light on the importance of treating issue writing as a first-class software engineering activity by investigating standardization around the writing of GitHub issues, and by developing novel solutions for issue writing, similar to decades of research on code writing. For instance, future research can investigate how to embed our predictive models directly into agent workflows so agents can proactively ask for missing and confusing information before starting the implementation (e.g., by using the theory of mind proposed by Gallaba et al.~\cite{gallaba2025towards}), and developing methods to split complex tasks into smaller units that better match current LLM capabilities.

We provide a replication package on GitHub~\footnote{Replication package: \url{https://github.com/DaREf-MS/pr_prediction_from_issues/}}.

The rest of the paper is organized as follows. Section~\ref{sec:background} discusses the necessary background to our paper. Section~\ref{sec:methodology} discusses the data collection for our paper and how we measure the quality criteria. Section~\ref{sec:preliminary_study} discusses the evaluation of our LLM solution in assessing the quality of issues based on the 32 criteria. 
Section~\ref{sec:results} discusses the results of our work. 
Section~\ref{sec:implications} discusses some implications of our work. Section~\ref{sec:related_work} discusses the closest work to our paper. Section~\ref{sec:threats} enumerates a set of threats to the validity of our results before concluding the paper in Section~\ref{sec:conclusion}.

\section{Background}
\label{sec:background}

The goal of this section is to discuss the process of GitHub issues, assigning the issue to an AI-agent, which takes the lead in implementing the task through a pull request. 

Practitioners can open a GitHub issue that describes the task to implement or the bug to fix, which will be assigned to an AI-agent such as Copilot. When one opens the GitHub issue, it can be discussed by developers to add relevant details such as the context about the feature, any indications on how to implement the task or fix the bug, and any other relevant information. When the discussion is over, developers can assign the GitHub issue to Copilot that opens a pull request and starts the implementation. The agent has the entire discussion in the GitHub issue up to the time when Copilot is assigned to the implementation as input. Once the implementation is done, the opened pull request can be reviewed to make the appropriate changes before merging it. 

To get better results from Copilot, GitHub provides a set of best practices for GitHub issues~\footnote{\url{https://docs.github.com/en/copilot/tutorials/coding-agent/get-the-best-results}} which also forms the basis for our study. For instance, the documentation mentions that the GitHub issue should have a clear description, complete acceptance criteria, context-related directions such as the files that need to be changed, and the tasks to implement should be well scoped, etc. 
Our goal in the paper is to concretize these best practices into measurable criteria quantified by LLMs, to study the extent to which these criteria matter, and to guide developers through a predictive model on the right aspects to improve in a GitHub issue before assigning it to an AI-agent such as Copilot.

\begin{table*}[]
\centering
\resizebox{\textwidth}{!}{
\begin{tabular}{
    p{0.3\linewidth}|
    p{0.5\linewidth}|
    c|c|c|c|c|c
}
\textbf{Category} & \textbf{Description} &
\rotatebox{90}{\textbf{Clear Definition}} &
\rotatebox{90}{\textbf{Acceptance Criteria}} &
\rotatebox{90}{\textbf{Directions (Files)}} &
\rotatebox{90}{\textbf{Scoped Task}} &
\rotatebox{90}{\textbf{Sensitive/Critical}} &
\rotatebox{90}{\textbf{Ambiguity}} \\ \hline

\multicolumn{8}{l}{\textbf{Problem Definition and Understanding (Prob.)}} \\ \hline
Clarity of the problem statement & Does the issue clearly describe what needs to be solved or implemented? & x &  &  &  &  &  \\ 
Problem comprehension & Does the issue provide enough background information (e.g., logs, stack traces, screenshots, or examples) to fully understand the problem? & x &  &  &  &  &  \\ 
Ambiguity avoidance & Does the issue avoid vague or open-ended phrasing, unclear goals, or underspecified requirements? &  &  &  &  &  & x \\ 
Contradiction avoidance & Does the issue avoid any contradicting statements or claims? &  &  &  &  &  & x \\ 
Scope of the task & Is the work small enough for one automated change and free of unrelated goals? &  &  &  & x &  &  \\ 
Detail and task-simplicity alignment & Does the issue provide an appropriate amount of detail relative to the task’s simplicity, including cases where no information is needed (e.g., dependency upgrade, adding a link, renaming a file, implementing another pull request)? & x &  &  & x &  &  \\ 
Expected and edge behaviors & Does it specify both nominal and exceptional cases to consider? &  & x &  &  &  &  \\ 
Readability and structure & Is the issue well organized (sections, bullets, markdown headings, clear formatting)? & x &  &  &  &  &  \\ 
Root cause articulation & Does the issue identify or hypothesize the underlying cause of the bug or limitation, not just the symptom? & x &  &  &  &  &  \\ 
Reproduction steps and evidence & Does the issue list concrete steps or minimal examples provided to reproduce or confirm the problem? & x &  &  &  &  &  \\ \hline

\multicolumn{8}{l}{\textbf{Acceptance and Validation Criteria (Acc/Val.)}} \\ \hline
Validation guidance & Does the issue describe how to verify correctness of the change (unit, integration, manual, end-to-end tests)? &  & x &  &  &  &  \\ 
Goal-test alignment & Do the success conditions logically align with the stated work to do (e.g., problem to fix or feature to implement)? & x & x &  &  &  &  \\ \hline

\multicolumn{8}{l}{\textbf{Implementation and Technical Context (Impl.)}} \\ \hline
Context guidance & Does the issue identify relevant files, modules, or components relevant to the work that the AI should do? &  &  & x &  &  &  \\ 
Solution direction guidance & Does the issue propose a plausible approach or direction without prescribing exact code? &  &  & x &  &  &  \\ 
Context setup / environment specification & Does the issue specify configuration, dependency, or version details relevant to the work to do? &  &  &  & x &  &  \\ 
Actionability granularity & Are the required actions broken down into manageable, concrete steps for implementation? &  &  &  & x &  &  \\ 
Common task familiarity & Does the task to implement refer to a common GitHub task for which AI-agents are faimilar with (e.g., configure dependabot, configure CI/CD, ...)? &  &  &  & x &  &  \\ 
System/component localization & Does the issue specify the affected subsystem or API explicitly (paths, classes, functions)? &  &  & x &  &  &  \\ 
Naviguation hints & Does the issue provide directly or indirectly hints that will help the AI-agent quickly navigate the project (e.g., parameters or function names easy to grep, a specific type of artifacts like tests, ...)? &  &  & x &  &  &  \\ 
Boundary and assumption clarity & Does the issue state environment assumptions, preconditions, or constraints (e.g., OS, framework version)? &  &  &  & x &  &  \\ 
Example relevance & Does the issue mention examples that are realistic and representative of real use cases? & x &  &  &  &  &  \\ 
Dependency awareness & Does the issue mention external APIs, libraries, or services it depends on or affects? &  &  & x &  &  &  \\ 
Refactor and migration awareness & Does the issue acknowledge code or API migrations that may affect implementation? &  &  &  & x &  &  \\ \hline

\multicolumn{8}{l}{\textbf{Risk and Quality Awareness (Risk)}} \\ \hline
Sensitive or critical task awareness & Does the issue identify any security, privacy, or safety implications, ensuring suitability for AI involvement? &  &  &  &  & x &  \\ 
Risk anticipation & Does the issue foresee possible side effects, regressions, or integration risks for which the AI will need human oversight to avoid harm? &  &  &  &  & x &  \\ 
Backward compatibility consideration & Does the issue mention how changes affect existing users or APIs? &  &  &  &  & x &  \\ 
Performance and scalability awareness & Does the issue mention performance constraints, latency, or efficiency targets? &  & x &  &  &  &  \\ 
Testing risk level & Does the issue describe potential testing difficulty (mocking, integration setup)? &  &   &  &  & x &  \\ \hline

\multicolumn{8}{l}{\textbf{Traceability and Completeness (Trace)}} \\ \hline
Traceability and linkage & Does the issue reference related issues, PRs, specs, or documentation for context directly through links or indirectly (e.g., reference to the open pull requests)? & x &  &  &  &  &  \\ 
Update recency and context freshness & Does the issue appear up-to-date (references current versions, not deprecated APIs)? & x &  &  &  &  &  \\ 
Self-containment & Can a developer or AI agent act on this issue using only the information provided? & x & x & x &  &  &  \\ 
Overall coherence and alignment & Do all parts of the issue (problem, hints, validation) logically connect? & x & x &  & x &  &  \\ 

\end{tabular}
}
\caption{Criteria for evaluating GitHub issues for AI-readiness. The six last columns represent the quality criteria that GitHub issues should consider according to GitHub Copilot documentation.}
\label{tab:rubrics}
\end{table*}

\section{Methodology}
\label{sec:methodology}


This section discusses the collection of GitHub issues and their associated pull requests in Section~\ref{sec:data_coll}, our set of criteria in Section~\ref{sec:rubrics}, and how we leverage GPT-5-mini to evaluate each GitHub issue based on our criteria in Section~\ref{sec:measure_rubrics}.

\subsection{Data Collection}
\label{sec:data_coll}

We leverage AIDev~\cite{li2025aidev} to collect pull requests created by an AI-agent, along with their associated GitHub issues. AIDev~\cite{li2025aidev} is a large scale dataset of pull requests created or co-authored by an AI-agent. 
We select from the benchmark pull requests that meet the following criteria: (1) The pull request is created by an AI-agent. We are interested in tasks that are implemented by an agent from natural language text (i.e., GitHub issues). Despite the fact that the AIDev dataset has five agents, the only agent that creates pull requests is Copilot, while the other agents are co-authoring commits without having their prompts available in the dataset or GitHub to study their quality. The reason for which we consider only Copilot. (2) The pull request has an associated GitHub issue. 

We select the associated GitHub issues, which we evaluate using our criteria. When developers assign a GitHub issue to an AI-agent, the agent will use the entire discussion available to create a pull request. Thus, a GitHub issue that we evaluate includes the title, the description, and all the comments up to the creation of the pull request by the AI-agent. Since the benchmark of AIDev~\cite{li2025aidev} does not contain the comments, we collect them using the GitHub API. 
We then filter out the comments to consider only these created before the pull request. Our dataset contains a final set of 3,180 GitHub pull requests and their associated issues.

\subsection{Evaluated Criteria}
\label{sec:rubrics}

Our goal is to define a set of criteria to assess the readiness of a GitHub issue for an AI-agent. To do so, we consider the criteria defined in Table~\ref{tab:rubrics}. Our criteria cover the clarity of the GitHub issue, how to accept and validate the implementation of a GitHub issue's task, implementation and context guidance, the link between GitHub issues and external artifacts, including other issues and specifications, as well as the definition of risks in GitHub issues as the final category of criteria. These criteria are defined based on our experience with GitHub issues, pull requests, and AI-agents, as well as the documentation of best practices for GitHub issues for GitHub Copilot~\footnote{\url{https://docs.GitHub.com/en/copilot/tutorials/coding-agent/get-the-best-results}}. The criteria and their wording were then improved and refined with GPT-5. 

The last six columns of Table~\ref{tab:rubrics} are the best practices suggested by Copilot, which are covered in a considerable amount of our criteria. Note that while the clear description, the acceptance criteria, and directions on which files to change are defined as best practices, the remaining three (i.e., scoped tasks, sensitive/critical tasks, and ambiguous tasks) are tasks that GitHub Copilot recommends developers implement themselves instead of using Copilot. We consider these aspects to see if developers are able to control the scope of their GitHub issues to make them easier for an AI-agent like Copilot, if GitHub issues with clear hints into the sensitive and critical aspects of the tasks would have a chance to be accepted, and whether developers manage to write tasks that are unambiguous. 

Each issue should have a score for each of the evaluated criteria that ranges between 0 and 5. 0 when the criterion is missing, 1 when it is poor, 2 when the criterion is weakly described, 3 when the criterioin is acceptable, and 4 and 5 when criterion is good and excellent, respectively. 

\subsection{Criteria' Measurements}
\label{sec:measure_rubrics}

We leverage an AI-agent (GPT-5-mini) and the self-consistency technique~\cite{wang2022self} to measure the quality of each issue according to the previously discussed criteria. Our prompt explicitly instructs the model to assess the readiness of the GitHub issue for the AI-agent. The system and user prompts are provided in the appendix. We configure GPT-5-mini with a ``\textit{high}'' reasoning effort. Following self-consistency, majority agreement is assumed to reduce prediction errors~\cite{wang2022self}. As such, we perform three independent runs of GPT-5-mini with the same configuration and aggregate predictions via majority voting.

\section{Preliminary Analysis: Reliability of LLM Criteria Evaluations}
\label{sec:preliminary_study}

The goal of this section is to evaluate our approach to assessing the quality of issues using GPT-5-mini with high reasoning effort. We perform such an evaluation through the following two preliminary research questions: 

\subsection*{PQ1. How consistent are LLM-based criterion evaluations of GitHub issues across repeated runs?}

\textbf{Motivation:} The goal of this preliminary analysis is to assess the consistency of GPT-5-mini when evaluating GitHub issues. Obtaining consistent evaluations, despite the high reasoning capabilities and configuration of GPT-5-mini, supports the reliability of our LLM-based evaluation approach.

\textbf{Approach:} To measure the robustness of our approach, we run GPT-5-mini three independent times with the same configuration discussed in the methodology Section on the studied issues. We then quantify the agreement score between the three agents using Krippendorff's alpha~\cite{krippendorff2018content}, which measures the degree to which repeated LLMs evaluations assign similar criterion scores (0 to 5) to the same issue. The alpha score is interpreted as follows: A score of 1 for perfect agreement score, $>$= 0.8 a satisfactory/strong agreement, $[0.67 - 0.79]$ moderate agreement, $<$ 0.67 poor agreement among raters, and a score of 0: no agreement, equivalent to a random rating. 

For disagreements, we further investigate the structure of the disagreement. We look for cases when only one LLMs diverges from the group and cases where all three LLMs disagree. That is inspired by the concept of self-consistency~\cite{wang2022self}, where majority runs are more likely to be reliable compared to individual predictions. Our investigation of the structure of disagreements also measures the magnitude of the disagreement, which is the difference between the maximum and minimum criterion scores produced across the three runs. A small difference indicates mild and acceptable differences, while large differences indicate unreliable scoring for that criterion. For example, predicting clarity as 1 or 2 reflects a poor-quality description and represents a minor discrepancy; however, predicting the same criterion as 1 and 5 indicates a high level of inconsistency and low reliability. This distinction is particularly important for subjective criteria, where small variations are still acceptable.

\textbf{Results: 23, 6, and 3 out of the 32 criteria show strong, moderate, and poor agreements among the three LLMs.} The highest agreement score is 93.49\%, while the lowest is 37.2\% for the criteria \textit{of traceability linkage} and \textit{sensitive tasks}, respectively. Traceability linkage can be easily quantified, as the LLM needs only to look for links and references to external artifacts. However, sensitive tasks require a better understanding of what is defined as sensitive and what is not for a given project, along with the contextual information to which the LLM does not have access. 

\textbf{Despite the disagreement, the difference between the three LLMs is a median of only 1 for all the disagreement cases}. Note that this median concerns only the issues with disagreement among the three LLM, where at least one LLM reports a different score than the other two. The average difference among the three LLMs (the difference between the maximum and minimum values for a given criterion) ranges between 1.02 and 1.75 for overall coherence and contradiction avoidance, respectively. We also observe that all the third quartiles are just one, with the exceptions of Root cause articulation, system localization, example relevance, and contradiction avoidance, whose Q3 values are respectively 2, 2, 2, and 3. Even for the \textit{sensitive tasks} criterion that shows the lowest agreement score, the disagreement difference is only one.

\textbf{85.43\% to 98.34\% of the disagreements are due to only one LLM having a different prediction than the other two.} In fact, the disagreements are rarely fully dispersed among the three models; rather, they are dominated by a single LLM behavior. This indicates that pair-wise agreement between the LLMs is frequent, and a simple vote labeling would reflect an agreement between at least two LLMs. This is also true for the \textit{sensitive task} criterion, which has the lowest agreement score. 95\% of the disagreement cases involve 2 LLM agreeing on a given score while one LLM has a different prediction.

\begin{Summary}{Summary of PQ1}{}
Our three LLMs are consistently scoring 71.8\% (23 out of 32) of the studied criteria with a strong agreement score. Despite the moderate and poor agreement scores for the other criteria, the disagreement is only +/- score of 1 and mostly have the shape of one of the three LLMs scoring differently a criterion. 
\end{Summary}

\subsection*{PQ2. How accurately can LLM distinguish low-quality from high-quality issues based on their criteria?}

\textbf{Motivation:} The goal of this RQ is to assess the quality of the LLM evaluations to identify the extent to which one can rely on our LLM pipeline to evaluate a given Github issue. 

\textbf{Approach:} Our evaluation consists of studying whether LLMs are able to distinguish high-scoring issues from poorly scored ones given each of our criteria. Since it is challenging to evaluate the correctness of the LLMs' scores at the granularity of the six scores (0 to 5), we evaluate whether LLMs can distinguish between high-quality and low-quality issues. We define high quality issues in a given criterion as those with a score of 4 (good) or 5 (excellent) on a studied criterion, while issues with a score of 0 to 3 are considered low quality in the same criterion. 

We then evaluate whether we can correctly identify the group of high quality issues from the group of low quality issues for each criterion. To do so, we generate 10 issues for each criterion, five of which are high-quality and the other five of which are low quality according to the LLM. Then, we manually investigate whether we can reach the same classification of the 10 issues in the two groups (high and low quality) blindly (without looking into the LLM prediction). Our manual analysis of the 32 criteria considers a total of 320 manually studied issues. 

In a second step, we evaluate whether the manual classification and the classification of the LLM are aligned. If not, we completely redo the cases with at least 4 misclassified instances. These 4 typically consist of 2 high-quality issues classified in the low-quality group, and the other 2 are low-quality issues misclassified as high-quality. Our reassessment leverages the exact same 10 issues without considering the predictions of the LLM and without reviewing the initial round classification to avoid any bias toward the predictions of the LLM or our first round of classifications. 

We evaluate the performance of the model using precision. It measures the total number of high-quality issues according to the LLM that are manually labeled as high-quality. For example, if the LLM classifies all the high-quality issues (according to our manual analysis) as high-quality issues, the precision is 100\%. The precision of the low-quality issues is exactly the same, as we consider a balanced dataset of equal size; if an issue is not manually classified as low-quality, it will be in the high-quality group. 

\textbf{Results: Considering the manual analysis as an oracle, the precision of the model is 88.75\%, which is high enough to rely on LLMs for evaluating Github issues}. From our first round of manually labeling the issues, we find a precision of 83.12\%. 132 issues labeled manually as bad issues are also labeled by the LLMs and self-consistency with a score between 0 and 3, whereas 28 are incorrectly predicted as good. Since we leverage an equal sample of 5 good and 5 bad issues, the prediction of good issues has the exact same distribution (132 vs. 28). The prediction of 12 criteria was 100\% correct; 20 criteria have at least one misprediction (one misclassified as good and consequently another one as bad). 13 out of the 20 cases involve one issue incorrectly predicted as high-quality and another issue predicted as low-quality. 6 out of the 20 cases have 4 issues misclassified, i.e., 2 issues incorrectly in the group of high-quality issues and two in the group of low-quality issues, whereas the other 6 issues are in the right group. One criterion has 6 issues misclassified. After fully reviewing the cases with at least 4 misclassifications, we obtain a precision of 88.75\%. Two criteria have two low-quality issues misclassified as good and vise versa. 16 criteria have at least one misclassification; 14 out of the 16 have only one low-quality issue in the high-quality group, and vise versa.

\begin{Summary}{Summary of PQ2}{}
Our pipeline of LLMs can accurately distinguish good from bad issues with a precision of 88.75\%, which suggest the reliability of our approach.
\end{Summary}

\section{Results}
\label{sec:results}

This section discusses the results of our empirical study on the association between issues and the merge rate of pull requests generated from the Github issues (RQ1), and evaluate our prediction model (RQ2) whose interpretation is studied in RQ3.

\textbf{RQ1. \rqii}

\textbf{Motivation:} The goal of this research question is to study the association between the quality of Github issues and the acceptance of their related pull requests. Our findings will shed light on whether there are factors that one should consider improving to increase the chances of the AI-agent generating pull requests that will eventually be merged. 

\textbf{Approach:} To answer our research question, we compare the merge ratio of pull requests generated from low quality issues and those generated from high quality issues. In particular, for each of our studied criteria, we measure the ratio of merged pull requests originating from low quality issues according to the studied criterion. A low quality issue according to a given criterion corresponds to a score of 0 (missing), 1 (poor), 2 (weak), or 3 (acceptable). We measure the merge ratio for each criterion for high-quality issues whose scores are 4 (good) or 5 (excellent). We then measure the difference between the two merge ratios, i.e., the merge rate of pull requests originating from high quality issues minus the merge rate of pull requests from low quality issues. A positive difference indicates that the merge rate is higher for pull requests generated from high quality issues and vise-versa. To make our comparison statistically sound, we leverage the Chi-square test with $\alpha$ = 0.05. 

\begin{figure}
    \centering
    \includegraphics[width=\linewidth]{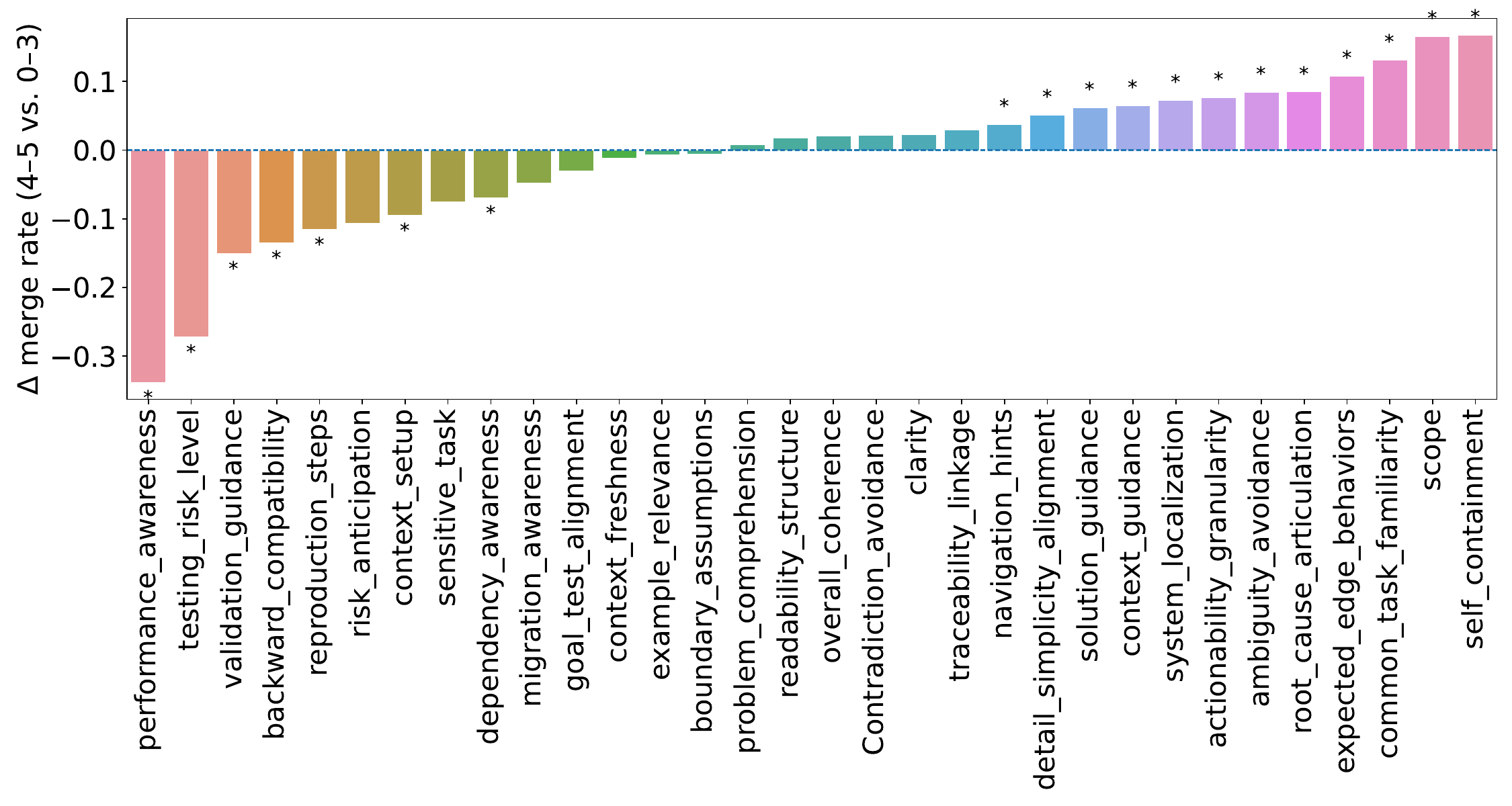}
    \caption{Merge rate difference between pull requests originated from high quality issues and low quality issues. * indicates that the difference is statistically significant (Chi-square test). }
    \label{fig:rubrics_eval}
\end{figure}

\textbf{Results: Well-scoped, self-contained issues that avoid ambiguities are associated with a higher merge rate}, as shown in Figure~\ref{fig:rubrics_eval}. Among all criteria, the scope of the task has one of the strongest positive associations with the merge rate. For instance, tasks whose scope is small enough for one automated change have a 16.44\% higher merge rate compared to issues with a low score (between 0 and 3). In addition to well-scoped tasks, tasks with a good balance between the amount of provided details and the simplicity of the tasks have a 5.08\% higher merge rate compared to tasks with a low score on the same criterion. The provided details should be explicit and avoid ambiguities (i.e., vague, open-ended phrasing, unclear goals, or underspecified requirements) and are associated with an 8.38\% higher merge rate compared to pull requests from issues with a low score for the same criterion. Within the topic of ambiguity and completeness of requirements, issues that score the best on detailing the expected and edge behaviors show a 10.69\% higher merge rate. In the same direction, having a well self-contained issue is associated with a 16.65\% higher merge rate. Despite the fact that detailed reproduction steps are strongly aligned with overall issue quality (a Spearman correlation of 72.56\%), we observe that issues with a higher score on reproduction steps are 11.53\% less likely to be merged compared to issues with a lower score. Note that the issues with high performance are 33.77\% less likely to be merged compared to those with the lowest performance awareness score. This requires further analysis by future work, as we observe a few issues (only 21) with a high performance awareness score. By examining all the pull requests with a performance awarness score of 4 and 5, we do not find a clear reason for their rejection.

\textbf{Providing guidance within the system without involving external configuration, dependencies, or APIs is also associated with higher merge rates.} In fact, issues with a good identification of relevant files, modules, or components for the task of the AI agent (\textit{context guidance}) and issues that more precisely specify the affected subsystem (\textit{System/component localization}) or provide good hints to help the AI agent navigate the project (\textit{navigation hints}), such as parameter or function names, are associated with 6.39\%, 7.22\%, and 3.68\% higher merge rates, respectively. However, a good description of external APIs or services that the issue depends on or affects (\textit{dependency awareness}) is associated with a 6.91\% lower merge rate. Once the files/components/modules are identified, providing good suggestions on how to implement the task (\textit{solution direction guidance}) is associated with a 6.14\% higher merge rate. As expected, tasks that are common for AI agents, such as configuring dependabot or CI/CD, are associated with a 13.10\% higher merge rate. Another way to provide guidance about how to implement the task is to further split it into manageable and concrete steps (\textit{Actionability granularity}), which is associated with a 7.58\% higher merge rate. This observation is consistent with the previous finding about the scope of the task. Similar to the previous observation, we observe that mentioning configuration, dependency, or version details relevant to the work to be done (\textit{Context setup/environment specification}) is associated with a 9.4\% lower merge rate.

By comparing the issues with a good score on all the criteria with a positive impact in Figure~\ref{fig:rubrics_eval}, i.e., from the criterion \textit{navigation hints} to \textit{self containment}, we find 118 issues that have high-scores for all those criteria and 292 that have low scores on all of the same criteria. We find that the merge rate is 31.2\% higher for the high-quality issues (a merge rate of 77.11\% compared to a merge rate of 45.91\%).

\begin{Summary}{Summary of RQ1}{}
Well-scoped, self-contained, and unambiguous issues are associated with a higher merge rate. 
Issues that provide clear guidance for locating relevant files, components, or parameters—including simple navigation hints—also show higher merge rates. 
Further, issues that offer concrete implementation guidance or break the task into manageable steps are associated with higher merge rates. 
In contrast, issues that emphasize external context—such as environment configuration, context setup, dependencies, or external APIs—are associated with lower merge rates.
\end{Summary}

\textbf{RQ2. \rqiii}

\textbf{Motivation: }The goal of this research question is to quantify how accurately we can predict the likelihood that a pull request will be merged based solely on its originating GitHub issue. This allows developers to improve issue quality before assigning it to Copilot, thereby increasing the chances that the resulting pull requests will be accepted.

\textbf{Approach: }We evaluate different machine learning algorithms, which include logistic regression, random forest, and XGBoost for our prediction. To train and test our models, we follow the approach of prior studies~\cite{Lee_2020, 10795090}. In particular, we consider the following steps: 

\begin{table}
    \centering
    \begin{tabular}{p{2.8cm}|p{5cm}}
        feature & Definition \\
        \hline
        num comments & The number of comments in the GitHub before the creation of the PR\\ 
        sum comment length & The textual length of the comments \\ 
        num unique commenters & Number of commenters on the issue before the creation of the PR\\ 
        issue body length & The length of the body of the issue\\ 
        issue title length & The length of the title of the issue\\
        num PRs before & Number of PRs created in the project before the studied PR \\
        num merged PRs before & The number of merged PRs before the studied PR\\
        num copilot PRs before &  The number of copilot PRs before  \\
         &  \\
    \end{tabular}
    \caption{Additional features about the overall context of the issue and the created pull request}
    \label{tab:general_features}
\end{table}

\textbf{Independent and dependent features:} Our independent features consist of the set of criteria defined in the methodology Section. Each GitHub issue is characterized by these criteria values, which range between 0 and 5, in addition to the criteria for the instruction files. In addition to the criteria, we add the features shown in Table~\ref{tab:general_features}, which pertain to the content of the issue and features about the project at the time of creating a pull request. We end up with a total of 40 features. The dependent variable in our model is the result of the pull request that implements the GitHub issue, which can be either merged or closed. Any open pull request is removed from our dataset. 

\textbf{Correlation and redundancy analysis:} As found by prior work~\cite{8608002}, correlated and redundant variables can negatively impact the interpretation of machine learning models. As such, we leverage Spearman correlation to remove one variable from each pair of correlated independent variables. We consider, similarly to prior work~\cite{Lee_2020, 10795090}, a correlation threshold of 0.7. We do not have any variables that are redundant in our dataset (i.e., a redundant variable is a variable that can be predicted using the other independent variables). We end up with a total of 21 independent variables.

\textbf{Training and testing:} Once the correlated and redundant variables are excluded, we train and test our machine learning models. For such training and testing, we leverage 100 bootstrap samples with replacement. We select a bootstrap sample of the same size as the whole data and with replacement for training a model, using any remaining data for testing. We repeat the same process of splitting data, training, and testing a model 100 times. For training a model, we evaluate different interpretable models, which are logistic regression, random forest, and XGBoost. For testing the model, we leverage the standard evaluation features: Area Under Curve (AUC), precision, and recall. AUC measures the discrimination ability of the model; the higher the AUC, the better the model is, with an AUC of 50\% representing a random guess. Precision measures the ability of the model to predict only positive cases (merged pull requests). Recall measures the ability of the model to predict merged pull requests as merged. The higher the precision and recall, the better the model is. 

To better understand whether all of our dimensions of criteria (i.e., problem definition, acceptance and validation criteria, implementation and technical context, risk and quality awareness, and traceability and completeness) have an impact on the prediction, we train and test a model with one dimension at a time, following the same training and testing process described earlier. Note that for training a model for one dimension, we consider all the features of the same dimension, as our goal is only to measure the performance of that dimension without interpretation; hence, there is no need to remove correlated features for dimension based models. If the model has an AUC higher than 50\% (i.e., random guess), we conclude that the dimension has explanatory power for predicting the likelihood of a pull request being merged from the dimension criteria only. We repeat the same process for each of the dimensions.

\begin{table}
    \centering
\scriptsize
\setlength{\tabcolsep}{3pt}
    \begin{tabular}{llccccccc}
    Model & feature & All & General & Acc/Val. & Impl. & Prob. & Risk & Trace \\
\midrule
Logistic & AUC & 0.651 & 0.609 & 0.561 & 0.608 & 0.638 & 0.533 & 0.568 \\
 Regression & Precision & 0.639 & 0.597 & 0.585 & 0.607 & 0.628 & 0.583 & 0.586 \\
 & Recall & 0.781 & 0.899 & 0.939 & 0.802 & 0.787 & 0.935 & 0.869 \\
\midrule
Random  & AUC & 0.720 & 0.688 & 0.543 & 0.589 & 0.583 & 0.522 & 0.549 \\
Forest & Precision & 0.691 & 0.686 & 0.582 & 0.621 & 0.613 & 0.582 & 0.591 \\
 & Recall & 0.774 & 0.714 & 0.925 & 0.687 & 0.659 & 0.859 & 0.695 \\
\midrule
XGBoost & AUC & 0.701 & 0.682 & 0.543 & 0.577 & 0.592 & 0.521 & 0.557 \\
 & Precision & 0.687 & 0.677 & 0.582 & 0.610 & 0.618 & 0.581 & 0.591 \\
 & Recall & 0.732 & 0.708 & 0.931 & 0.657 & 0.686 & 0.872 & 0.704 \\
\midrule
    \end{tabular}
    \caption{Performance of our evaluated prediction models}
    \label{tab:performances}
\end{table}

\textbf{Results: Random forest shows the best discrimination ability with a median AUC of 72\%, followed by XGBoost with a median AUC of 70.1\%}, as shown in Table~\ref{tab:performances}. Logistic regression has the lowest median AUC of only 65.1\%. We observe that the models are consistent over the 100 iterations, with an AUC variance of 2.92x$10^{-4}$ and 3.20x$10^{-4}$ for Random Forest and XGBoost, respectively. Our Random Forest and XGBoost models show a precision of 69.1\% and 68.7\% and a recall of 77.4\% and 73.2\%, respectively. This indicates that the features we used are good predictors for the acceptance of the pull requests and provide stable models. Note that we do not expect our model to be 100\% accurate, as other factors might interfere with the acceptance of a pull request even when the generated code is correct (e.g., the feature is not a priority).

\textbf{All of our studied dimensions have enough explanatory power (AUC higher than 50\%) to predict the likelihood of a pull request being merged,} as shown in Table~\ref{tab:performances}. General features that define the project and the amount of text in the issue have higher prediction capabilities compared to the other dimensions (whose impact will be further discussed in the next RQ). Providing implementation details and technical context is the second dimension that predicts the likelihood of a pull request being accepted, slightly higher than the definition of the problem. External traceability and risk related criteria are the lowest dimensions with explanatory power for distinguishing GitHub issues that lead to an accepted from a rejected pull request. We also observe that the model with all the dimensions has better performance. Note that we remove correlated features from the model with all the features, as we are interested in building an interpretable machine learning model that can also guide practitioners on how to improve their issues for getting an accepted pull request.

\begin{Summary}{Summary of RQ2}{}
Our random forest model is able to predict the chances of a pull request to be merged from the GitHub issue with a median AUC, precision, and recall of 72\%, 69.1\%, and 77.4\%, respectively. All of our dimensions have an explanatory power higher than random guess, whereas the combination of all dimensions showing better performances. 
\end{Summary}

\textbf{RQ3. \rqiv}

\textbf{Motivation:} The goal of this research question is to understand which features drive the prediction model. Identifying these features can help practitioners focus on what to improve to increase the likelihood of a merged pull request. 

\textbf{Approach:} To identify the most important features, we leverage the random forest model that demonstrates the best predictive performance in the previous research question. 

\textit{- Ranking of the most important features.}To identify the most important features, we use the internal feature importance of the random forest combined with scott-knott clustering. From each of the 100 bootstrap iterations, we extract feature importance scores, which yields to 
100 featue rankings, which we aggregate using Scott-knott~\cite{tantithamthavorn2017mvt, tantithamthavorn2018optimization} into a final ranked list of features.

\textit{- Impact of features on the probability of an issue ending up with a merged pull request}. We quantify how changes in individual features (e.g., increasing the \textit{scope} score) affect the predicted probability. 
To do so, we leverage \textit{Partial Dependence}~\footnote{\url{https://christophm.github.io/interpretable-ml-book/pdp.html}}, which measures how the predicted probability changes with the variation of one feature at a time. We also study pairwise feature interactions using . 
\textit{Partial Dependence} with two features at a time. 

For each criterion, we quantify three metrics: (1) the effect type, (2) the direction of the impact, and (3) the magnitude of the impact. \textbf{Effect}: Based on the partial dependence values, we fit a linear model (y=ax+b) to capture the association between a feature and the predicted probability. 
The relation is considered linear if $R^2$ of the model is $> 98\%$ and Spearman correlation $> 75\%$; monotonic if only the correlation $> 75\%$; Otherwise, the relation is a complex one. \textbf{Direction}: We identify whether the association bewteen a feature and the predicted value is positive or negative, 
we examine the sign of the linear coefficient (the coefficient a of y = ax + b). 
\textbf{Amplitude}: We measure the magnitude of impact as the difference between the maximum and minimum predicted probabilities observed when varying a feature.

\begin{table}
    \centering
    \begin{tabular}{p{0.3cm}p{3.1cm}p{1cm}p{0.4cm}p{0.4cm}p{0.4cm}p{0.4cm}}
\toprule
rank & feature & effect & direc. & r2 & Spear. & amp. \\
\midrule
1 & issue body length & monotonic & $\searrow$ & 0.41 & 0.81 & 0.09 \\
2 & issue title length &  complex & $\nearrow$ & 0.18 & 0.21 & 0.15 \\
3 & num merged prs before  & monotonic & $\searrow$ & 0.73 & 0.87 & 0.12 \\
4 & sum comment length  & monotonic & $\searrow$ & 0.79 & 0.88 & 0.11 \\
5 & num copilot prs before  & complex & $\nearrow$ & 0.38 & 0.64 & 0.07 \\
5 & scope  & monotonic & $\nearrow$ & 0.91 & 1.00 & 0.13 \\
6 & context guidance  & complex & $\nearrow$ & 0.56 & 0.71 & 0.03 \\
7 & root cause articulation  & monotonic & $\nearrow$ & 0.75 & 0.94 & 0.04 \\
8 & solution guidance  & monotonic & $\nearrow$ & 0.87 & 0.89 & 0.02 \\
9 & dependency awareness & monotonic & $\searrow$ & 0.95 & 1.00 & 0.05 \\
9 & traceability linkage & monotonic & $\searrow$ & 0.78 & 0.83 & 0.02 \\
10 & example relevance  & complex & $\searrow$ & 0.03 & 0.03 & 0.01 \\
11 & context setup  & monotonic & $\searrow$ & 0.87 & 0.94 & 0.04 \\
12 & context freshness  & complex & $\searrow$ & 0.16 & 0.37 & 0.01 \\
13 & common task familiarity  & monotonic & $\nearrow$ & 0.80 & 0.94 & 0.07 \\
14 & overall coherence  & monotonic & $\nearrow$ & 0.82 & 0.90 & 0.02 \\
14 & migration awareness  & monotonic & $\searrow$ & 0.92 & 1.00 & 0.02 \\
15 & risk anticipation  & monotonic & $\searrow$ & 0.90 & 0.94 & 0.03 \\
16 & sensitive task  & complex & $\nearrow$ & 0.60 & 0.43 & 0.01 \\
17 & performance awareness  & monotonic & $\searrow$ & 0.86 & 1.00 & 0.05 \\
18 & Contradiction avoidance & complex & $\nearrow$ & 0.58 & 0.60 & 0.01 \\
\bottomrule
\end{tabular}

    \caption{The impact of each feature on the prediction probability. direc., Spear., and amp. stand for Direction, Separman correlation, and amplitude.}
    \label{tab:most_important_features}
\end{table}

\textbf{Results: Consistent with the high performance of the general category (the findings of the previous research question), the top features pertain to the general category of features}, as shown in Table~\ref{tab:most_important_features}. These features cohesively suggest that the description should be shorter without delving into the exact content of the issues addressed by the other features. In fact, increasing the body length or the number of comments can decrease the likelihood of a pull request being merged by 9\% and 11\%, respectively. While the title of an issue has a complex association with the probability of its pull request being merged, we observe that increasing the title length to up to 20 words can elevate the probability of a pull request being merged from 47\% to 60\%. After 20 words, the probability becomes stable and ranges between 0.58 and 0.62. 

\textbf{The content of the issue should target a well defined task small enough for one automated change and free from unrelated goals}. As shown in Table~\ref{tab:most_important_features}, restricting the scope of an issue is the fifth most important feature and can increase the chances of a pull request being accepted by 13\%. Root cause articulation (i.e., \textit{``Does the issue identify or hypothesize the underlying cause of the bug or limitation, not just the symptom?''}) when better defined can increase the chances of a pull request being accepted by 4\%. This can be explained by the fact that our data and model are not dedicated solely to bugs but to all types of GitHub issues. This can also be explained by the interaction of features. For instance, increasing the root cause articulation in a long description of 3502 words increases the probability from 53.8\% to 57.7\%; decreasing the length of the description from 3502 words to 596 words increases the probability from 53.8\% to 57.8\%. Doing both, i.e., increasing the root cause articulation from 0 to 5 and reducing the length of the description from 3502 words to 596 words, increases the probability from 53.8\% to 61.8\% (according to our partial dependency analysis of interacting features). Avoiding contradiction can also increase the chances of a pull request being merged, yet by only 0.1, which can be explained by the fact that most of our data (97.7\%) has a contradiction avoidance score of 5. The issue should be cohesive, as referencing external issues, pull requests, specifications, or documentation has a negative (a decrease of 2\%) association with the probability of a pull request being merged.

\textbf{Issues with better solution guidance or those related to familiar tasks and without involving third party dependencies are more likely to be merged.} In fact, issues that better propose solutions as approaches or directions without the exact code location (\textit{Solution guidance}) can increase the chances of the pull request being accepted by 2\%. GitHub issues about familiar tasks (\textit{common task familiarity}) are also more likely to be merged, with a probability increase of 7\%. We also observe a cohesive conclusion from three criteria related to the external context of a project. criteria related to external contexts, such as infrastructure configuration or third party libraries, components, or dependencies, are negatively associated with the probability of a pull request being merged. These features include dependency awareness, context setup, and migration awareness, which respectively decrease the merge probability by 5\%, 4\%, and 2\%. This can be explained by the fact that developers emphasize the external context for tasks that require such a context, which are less likely to be trivial for an AI-agent. Further studies are required in this direction to see how to better incorporate external contexts for AI-agents. In fact, 26\%, 45\%, and 64\% of the issues are missing dependency awareness, context setup, and migration awareness. Even better referencing up-to-date dependencies (i.e., context freshness) decreases the chances of a pull request being merged by 1\%.

\textbf{Risks related features are among the least important features and are associated with a lower prediction probability.} Despite being ranked at a late position (15, 16, and 17), the risk related criteria have a negative impact of 2\% to 5\% on the likelihood of a pull request being merged. The sensitive task has a complex association with the probability of a pull request being merged. This can indicate, as suggested by GitHub Copilot, that risky tasks should not be assigned to agents even when anticipating risks or performance impacts. Another explanation is that more than half of the studied issues are missing risk anticipations, and only 8.3\% of all the studied issues have a score equal to or higher than 3 for the same criterion. Performance awareness is lacking for 92.8\% of the studied issues, and less than 2\% have a score equal to or higher than 3.

\textbf{We also observe that the higher the number of merged pull requests, the less likely a Copilot pull request will be merged}. As shown in Table~\ref{tab:most_important_features}, the number of merged pull requests is ranked within the top 3 most important features and can have a reduction of 12\% on the probability of an issue ending up with a merged pull request. Such an observation on the amount of merged changes in a given project can reflect its size. A project with a higher number of merged pull requests is usually expected to be a larger project; hence, it is more difficult to change by agents. The number of prior copilot pull requests, although ranked fifth, has a complex association with the probability of a pull request being merged.

\begin{Summary}{Summary of RQ3}{}
Developers should make their pull request shorter and well scoped, provide guidance on the implementation to increase their chances of the pull request to be merged. The context in which Copilot is used needs to have less merged pull requests (smaller projects). Referencing external artifacts or anticipating risks is associated with a lower probability of pull requests to be merged suggesting further studies on how to better incorporate such contexts. 
\end{Summary}

\section{Implications}
\label{sec:implications}

This section discusses some implications for improving the integration of agents by developers and guides future research on how to enhance the use of agents with the goal of better integrating them as teammates. 

\subsection{Implications for Developers}

\textbf{We recommend that developers reduce and control the scope of their tasks}. We find in our study that reducing the scope is associated with a higher merge rate, and it is a good predictor of the likelihood of resulting in a merged pull request. In the same vein, asking agents to develop simple tasks for which LLMs are familiar is more likely to result in a merged agentic pull request. Further splitting the task into concrete and manageable steps increases the likelihood of resulting in a merged pull request. Accordingly, developers need to define the right granularity for their tasks.

\textbf{We suggest developers provide clear instructions on how to implement the solution.} Providing guidance on the solution, the root cause to help the agent understand what to fix, and context helps models navigate the project and implement the right fixes, which results in merged pull requests. Issues need to be self-contained so that agents do not have to navigate other artifacts to better understand the requirements or the solution to implement. While we recommend issues to be self-contained, we do not recommend developers create verbose issues. The shorter the description, the more likely the issue results in a merged pull request. 

\textbf{We suggest developers use our model to predict the likelihood of their issue resulting in a merged pull request.} We suggest developers use our model to measure the quality of their issue and identify what to improve before assigning the issue to Copilot. We opt for an interpretable machine learning model so that users can identify, using different interpretation techniques, the right metrics to improve in order to increase the probability of their issue resulting in a merged pull request. 

\textbf{Our finding hints at the fact that large projects are less likely to benefit from AI} agents. We find that the higher the number of pull requests, which can be considered a proxy for the amount of previously integrated changes and the size of the project, the less likely they are to have an issue resulting in a merged pull request. As such, we believe that reducing the size of the project by improving its modularity will further enhance the integration of AI-agents as coding teammates. The need to migrate to a modular architecture (e.g., microservices) is becoming increasingly important for the future of integrating AI-agents into software development activities. 

\subsection{Implications for Researchers}

\textbf{We recommend further research on issue writing as a first-class software engineering activity in the new era of AI-teammates}. As a follow up to a significant amount of software engineering research that aimed for decades to improve the quality of code writing, we shed light in this paper on the importance of writing GitHub issues. Our paper starts by proposing a set of criteria to help developers write GitHub issues, while further studies are needed to standardize such a process and support it with tools.

\textbf{We suggest that researchers investigate how to handle external contexts in GitHub issues.} We find that external contexts are associated with lower merge rates and have a negative impact on the predicted probability. Looking into how to better integrate external information about a GitHub issue can help agentic AI generate pull requests that end up being merged. For instance, future studies can examine the relevant information from external APIs and dependencies that are necessary for a given GitHub issue. An example of this information could be the implementation of the required APIs, the documentation of external dependencies, etc. 

\textbf{We suggest that researchers develop strategies that better integrate our machine learning model.} While we recommend developers use our model to improve GitHub issues, future studies can integrate our model as an agent/tool that can be leveraged directly by the AI-agent to ask users for the right information before starting the implementation. This can be part of clarifying users' intent, proposing the theory of mind as suggested by prior work~\cite{gallaba2025towards}.

\textbf{We suggest that future studies investigate tasks that are less likely to be merged. }For instance, we observe that tasks that effectively describe performance issues are rare and less likely to be merged. The same is true for tasks that involve testing difficulties, which have a lower merge rate. This can be explained by the lack of tools provided to AI-agents, such as tools to measure the appropriate performances and tools to run advanced tests (e.g., mocking). 

\textbf{We suggest that researchers develop approaches that split complex and large tasks into smaller units.} 
We recommend that future studies develop approaches that migrate complex tasks for LLMs into manageable small tasks on which LLMs can be more efficient. In fact, the scope is among the most important features for the model and is associated with a higher merge rate. Reducing the scope is, therefore, a task that can be further improved by future studies.

\section{Related Work}
\label{sec:related_work}

The closest work to our study corresponds to the evaluations of different prompt techniques for software engineering tasks (Section~\ref{sec:prompts_related_work}) and the work that studied the pull request mechanism (Section~\ref{sec:pull_requests_rw}).

\subsection{Prompt engineering for software engineering}
\label{sec:prompts_related_work}

The closest work to our paper are studies that focus on how to optimize prompts for software development.

Different prompt engineering techniques have been proposed, such as zero-shot~\cite{xian2017zero}, few-shot learning~\cite{brown2020language}, chain-of-thought~\cite{wei2022chain}, self-consistency~\cite{wang2022self}, and Tree-of-thought~\cite{long2023large}, and evaluated across programming aspects. White et al.~\cite{white2023prompt} proposed a catalog of prompt patterns. Siddiq et al.~\cite{10795375} studied the quality of prompts from code generation benchmarks such as HumanEval~\cite{chen2021evaluating}. They found that prompts suffer from formatting issues, misaligned sentences with the problems, and unnecessary tokens. Porta et al.~\cite{della2025prompt} studied the relation between prompt engineering techniques and code quality by using the benchmark of Dev-GPT~\cite{xiao2024devgpt}. They found that the most common prompt engineering technique used by developers is zero-shot and that there are no significant differences among the prompt engineering techniques in terms of code quality, defined by maintainability, reliability, and security issues. Tony et al.~\cite{10.1145/3722108} studied different prompt engineering techniques and their impact on the security of the code generated by LLMs. They found that the Recursive Criticism and Improvement (RCI) technique~\cite{NEURIPS2023_7cc1005e} reduces security weaknesses across their evaluated LLMs (GPT-3, GPT-3.5, and GPT-4). Khojah et al.~\cite{khojah2025impact} studied the impact of different prompt engineering techniques (e.g., few-shot~\cite{brown2020language}, persona~\cite{tseng2024two}, and chain-of-thought~\cite{wei2022chain}) on function level code generation. They found that while some prompt techniques influence the generated code, combining different techniques does not help. The prompt techniques present a trade-off between quality and correctness. In the same direction, Fagadau et al.~\cite{fagadau2024analyzing} studied the impact of eight prompt features related to the style and content of the prompt on the correctness, complexity, size, and correctness of the generated methods by co-pilot. They concluded that an effective prompt includes a summary of the method to implement, a description of its behavior, input-output examples, and the use of the present-tense. However, contextual information, including boundary cases, has little impact.

A recent study~\cite{10.1145/3771933} found that the reasoning capabilities of advanced models reduce the need for sophisticated prompt engineering techniques. Those techniques designed for earlier LLMs could negatively impact the performance of advanced models. For such models, prompts require high-quality instructions and guidance rather than strict adherence to a prompt engineering technique. In a same direction, we leverage in a recent work~\cite{gallaba2025towards} the theory of mind concept to clarify users' intentions in order to build a step-by-step workflow for implementing those intentions. 

Our work builds on this foundation by identifying quality-related criteria and their likelihood of leading to a merged pull request when implemented by an AI-agent.

\subsection{Pull Requests}
\label{sec:pull_requests_rw}

A large body of research~\cite{10795090, 10.1145/2568225.2568260, 7194588, 20155201729664, 10.1145/3479529, 10.1145/3597208, hasan2023understanding, Wessel_2022, 10.1145/3540250.3549081, 10.1145/3540250.3549119} has extensively studied pull requests from different perspectives, including the events that occur during the pull request review process, the results of the pull request (being accepted or rejected), and the impact of a given pull request after merging. Gousios~et~al.~\cite{10.1145/2568225.2568260} studied factors that are associated with the acceptance of PRs and the acceptance time. Gousios~et~al.~\cite{7194588} found that maintainers decide to accept a PR based on its quality, adherence to the project's style and architecture, and test coverage. Other studies focused on different types of events that occur in pull requests, such as the @-mention~\cite{20155201729664}, referencing~\cite{10.1145/3479529}, and Github reactions~\cite{10.1145/3597208}. 
Bots, when properly configured during code review, can reduce the time to receive the first response, according to Hasan~et~al.~\cite{hasan2023understanding}. Wessel et al. ~\cite{Wessel_2022} also found that bots increase the number of merged pull requests. On top of bots, other studies~\cite{10.1145/3540250.3549081, 10.1145/3540250.3549119} proposed approaches to automatically generate comments. Finally, other studies such as the work of Mcintosh et al.~\cite{mcintosh2016emse} focused on predicting the impact of code review practices on post-release defects.

While these studies focused on human-generated pull requests, our goal is to investigate AI-generated pull requests and their chances of being merged or closed. Our work complements this line of research, particularly the work that predicts the chances of a pull request being merged by looking into improving the quality of Github issues to increase the likelihood of a pull request being merged.

\section{Threats to Validity}
\label{sec:threats}

The first threat to validity concerns the generalizability of our approach to other AI-agents, aside from GitHub Copilot. We do not generalize our results to other agents like Claude due to a limitation in the dataset at the stage of writing the paper. AIDev pull requests that are linked to GitHub issues are all pull requests made with AI-agents. That said, GitHub Copilot is one of the leading AI-agents in software programming. Future studies are needed to replicate our work for other AI-agents. 


An internal threat to the validity of the results concerns the 32 criteria that we propose. These criteria do not represent all the quality checks for a prompt; other criteria can still be defined. In addition, measuring the criteria using LLMs can introduce some inaccuracies. That said, we evaluate LLMs to see if the results are consistent and accurate. Even if some criteria are inconsistent, the differences among LLMs are as small as a difference of +/- one and occur when two agents already agree on the score. That can be solved by self-consistency~\cite{wang2022self}, which we consider a mitigation for potential inaccuracies in the data.

\section{Conclusions and Future Work}
\label{sec:conclusion}

AI-agents are being widely adopted and extensively evaluated for different programming tasks to the point that they are now incorporated as team collaborators. One can define a task to implement or a bug to fix in a GitHub issue, which is assigned to an AI-agent such as Copilot that performs the task. However, as suggested by GitHub Copilot, the AI-agent performs better with high quality prompts, i.e., the description and comments on a GitHub issue before being assigned to the AI-agent. 

In this paper, we evaluate the association between the quality of GitHub issues, defined using 32 criteria, and the likelihood of the AI-agent producing an accepted implementation of the GitHub issue. Our 32 criteria cover five dimensions, including the definition of the problem, the criteria that the agent can use to accept or validate its changes, the implementation and technical context, links to other content, and whether any risks are anticipated. We find that issues that are well scoped, self-contained, and unambiguous provide clear guidance about what to change; details about the implementation are associated with a higher merge rate. However, issues with external contexts like configuration, APIs, dependencies, and context setup are associated with a lower merge rate. To help developers improve their issues before assigning them to AI-agents, we leverage an interpretable machine learning model that, based on the Github issue, predicts the likelihood of the generated pull request being merged. The model achieves a high AUC of 72\%. The interpretation of our model is consistent with empirical observations; we find that shorter descriptions with well scoped tasks and guidance on the implementation increase the predicted probability of a pull request being merged. Referencing external artifacts decreases this probability. 

In future work, we plan to further improve our model by using new quality criteria and integrating our model as part of the AI-agents process to request any uncertainties or lack of relevant information from developers before starting the task. 

\bibliographystyle{IEEEtran}
\bibliography{bibfile}

\newpage

\onecolumn   
\setcounter{section}{0}

\section{Appendix}

The following is the system prompt for our AI-agent to quantify the quality of an issue:

\begin{tcolorbox}[promptstyle,breakable]
\begin{MyVerbatim}
You are an expert software engineering evaluator specialized in assessing the extent to which Github issues are suitable inputs for **AI-based autonomous** code generation (e.g., GitHub Copilot, GPT-based coding systems).

Your goal is to evaluate how well a given GitHub issue can serve as a direct, unambiguous, and actionable specification for autonomous code generation by AI agents (e.g., GitHub Copilot, GPT-based coding systems).

You must:
1. Read the Github issue title, description, and comments carefully.
2. Evaluate it using the provided rubric criteria, which measure:
- Problem clarity, scope, and contextual completeness
- Presence of validation and acceptance criteria
- Implementation guidance (files, components, constraints)
- Risk and dependency awareness
3. Assign a score from 0 to 5 for each criterion (0 = missing, 5 = excellent).
4. Output your evaluation strictly in JSON, including an overall score, recommendation, and short textual feedback.

You are objective, concise, and analytical — not conversational.
Avoid speculation or opinion; rely solely on the issue title, description, and comments and your rubric expertise.
\end{MyVerbatim}
\end{tcolorbox}

The following is the user prompt for our AI-agent to quantify the quality of an issue: 

\begin{tcolorbox}[promptstyle,breakable]
\begin{MyVerbatim}
Your goal is to assess **how directly actionable** the given GitHub issue is for an **AI code generation agent** such as GitHub Copilot.Focus on **instructional clarity** rather than human project management detail. The ideal issue is one that an AI model could:
- **Understand what code to write** (input/output, function, API, or behavior).
- **Identify where to modify** (file names, classes, or functions).
- **Implement without new design decisions** (clear acceptance criteria, edge cases, and examples).
- **Validate its output** (defined tests, expected results).

High scores reflect AI-readiness — concise, specific, and executable descriptions that minimize ambiguity. Low scores reflect issues that are too abstract, exploratory, or context-heavy, which will cause the AI to misunderstand the task to do, hallucinate, speculate, or perform incomplete tasks. 

# Task

Based on the given Github issue, evaluate whether an AI agent can implement and validate the task without external clarification or undocumented assumptions? Your evaluation should be according to the following rubrics. For each criterion, assign a score between **0 and 5**:
**0 = missing, 1 = poor, 2 = weak, 3 = acceptable, 4 = good, 5 = excellent.**


# Rubrics:

The following rubrics measure how well the issue functions as a natural-language specification for automated code generation, not how detailed or formal it is for human review.

## Problem Definition and Understanding

1. **Clarity of the problem statement** – Does the issue clearly describe what needs to be solved or implemented?  
2. **Problem comprehension** – Does the issue provide enough background information (e.g., logs, stack traces, screenshots, or examples) to fully understand the problem?  
3. **Ambiguity avoidance** – Does the issue avoid vague or open-ended phrasing, unclear goals, or underspecified requirements?  
4. **Contradiction avoidance** - Does the issue avoid any contradicting statements or claims?
5. **Scope of the task** – Is the work small enough for one automated change and free of unrelated goals?  
6. **Detail and task-simplicity alignment** - Does the issue provide an appropriate amount of detail relative to the task’s simplicity, including cases where no information is needed (e.g., dependency upgrade, adding a link, renaming a file, implementing another pull request)?
7. **Expected and edge behaviors** – Does it specify both nominal and exceptional cases to consider?  
8. **Readability and structure** – Is the issue well organized (sections, bullets, markdown headings, clear formatting)?  
9. **Root cause articulation** – Does the issue identify or hypothesize the underlying cause of the bug or limitation, not just the symptom?  
10. **Reproduction steps and evidence** – Does the issue list concrete steps or minimal examples provided to reproduce or confirm the problem?


## Acceptance and Validation Criteria

11. **Validation guidance** – Does the issue describe how to verify correctness of the change (unit, integration, manual, end-to-end tests)?  
12. **Goal-test alignment** – Do the success conditions logically align with the stated work to do (e.g., problem to fix or feature to implement)? 


## Implementation and Technical Context

13. **Context guidance** – Does the issue identify relevant files, modules, or components relevant to the work that the AI should do?
14. **Solution direction guidance** – Does the issue propose a plausible approach or direction without prescribing exact code?  
15. **Context setup / environment specification** – Does the issue specify configuration, dependency, or version details relevant to the work to do?  
16. **Actionability granularity** – Are the required actions broken down into manageable, concrete steps for implementation?  
17. **Common task familiarity** - Does the task to implement refer to a common Github task for which AI-agents are faimilar with (e.g., configure dependabot, configure CI/CD, ...)?  
18. **System/component localization** – Does the issue specify the affected subsystem or API explicitly (paths, classes, functions)?  
19. **Naviguation hints** - Does the issue provide directly or indirectly hints that will help the AI-agent quickly navigate the project (e.g., parameters or function names easy to grep, a specific type of artifacts like tests, ...)?
20. **Boundary and assumption clarity** – Does the issue state environment assumptions, preconditions, or constraints (e.g., OS, framework version)?  
21. **Example relevance** – Does the issue mention examples that are realistic and representative of real use cases?  
22. **Dependency awareness** – Does the issue mention external APIs, libraries, or services it depends on or affects?  
23. **Refactor and migration awareness** – Does the issue acknowledge code or API migrations that may affect implementation?


## Risk and Quality Awareness

24. **Sensitive or critical task awareness** – Does the issue identify any security, privacy, or safety implications, ensuring suitability for AI involvement?  
25. **Risk anticipation** – Does the issue foresee possible side effects, regressions, or integration risks for which the AI will need human oversight to avoid harm?  
26. **Backward compatibility consideration** – Does the issue mention how changes affect existing users or APIs?  
27. **Performance and scalability awareness** – Does the issue mention performance constraints, latency, or efficiency targets?  
28. **Testing risk level** – Does the issue describe potential testing difficulty (mocking, integration setup)?


## Traceability and Completeness

29. **Traceability and linkage** – Does the issue reference related issues, PRs, specs, or documentation for context directly through links or indirectly (e.g., reference to the open pull requests)? 
30. **Update recency and context freshness** – Does the issue appear up-to-date (references current versions, not deprecated APIs)?  
31. **Self-containment** – Can a developer or AI agent act on this issue using only the information provided?  
32. **Overall coherence and alignment** – Do all parts of the issue (problem, hints, validation) logically connect?


# Constraints:
1. Your evaluation should reflect whether the issue provides enough structured, unambiguous guidance for **an autonomous coding agent (e.g., Github co-pilot or a similar coding agent)** to produce a working implementation without further clarification.
2. Do not reward excessive narrative or discussion; prioritize specificity, structure, and directly usable technical detail.
3. Output Format (**strictly adhere**)

```json
{{
  "clarity": 0-5,
  "problem_comprehension": 0-5,
  "ambiguity_avoidance": 0-5,
  "Contradiction_avoidance": 0-5,
  "scope": 0-5,
  "detail_simplicity_alignment": 0-5,
  "expected_edge_behaviors": 0-5,
  "readability_structure": 0-5,
  "root_cause_articulation": 0-5,
  "reproduction_steps": 0-5,
  "validation_guidance": 0-5,
  "goal_test_alignment": 0-5,
  "context_guidance": 0-5,
  "solution_guidance": 0-5,
  "context_setup": 0-5,
  "actionability_granularity": 0-5,
  "common_task_familiarity": 0-5,
  "system_localization": 0-5,
  "navigation_hints": 0-5,
  "boundary_assumptions": 0-5,
  "example_relevance": 0-5,
  "dependency_awareness": 0-5,
  "migration_awareness": 0-5,
  "sensitive_task": 0-5,
  "risk_anticipation": 0-5,
  "backward_compatibility": 0-5,
  "performance_awareness": 0-5,
  "testing_risk_level": 0-5,
  "traceability_linkage": 0-5,
  "context_freshness": 0-5,
  "self_containment": 0-5,
  "overall_coherence": 0-5,
  "overall_score": "XX/160",
  "recommendation": "AI-agent-ready | Needs revision | Not ready",
  "summary_feedback": "A concise paragraph summarizing key strengths and weaknesses of the issue."
}}
```

# Input: 

<Issue title>
{title}
</Issue title>

<Issue description> 
{description}
</Issue description>

<Issue comments> 
{issue_comments}
</Issue comments>

\end{MyVerbatim}
\end{tcolorbox}

\end{document}